\documentclass[12pt,a4paper]{article}
\usepackage{amssymb,latexsym}
\usepackage{amsmath}
\usepackage[OT2,OT1]{fontenc}
\def\cyr{\fontencoding{OT2}\fontfamily{wncyr}\selectfont}
\usepackage{url}
\usepackage{hyperref}

\newcounter{mnotecount}[section]
\renewcommand{\themnotecount}{\thesection.\arabic{mnotecount}}
\newcommand{\mnote}[1]
{\protect{\stepcounter{mnotecount}}$^{\mbox{\footnotesize  $
      \bullet$\themnotecount}}$ \marginpar{\raggedright\tiny
    $\!\!\!\!\!\!\,\bullet$\themnotecount: #1} }

\DeclareMathOperator{\ch}{\cosh}
\DeclareMathOperator{\sh}{\sinh}
\DeclareMathOperator{\arsh}{arsinh}

\newcommand{\tPhi}{\widetilde{\Phi}}
\newcommand{\tf}{\widetilde{f}}
\newcommand{\tu}{\widetilde{u}}

\newcommand{\sqhwi}[1]{\sqrt{\left.\det \hdw\right|_{#1}}}
\newcommand{\sqhzwi}[1]{\sqrt{\left.\det \hdzw\right|_{#1}}}
\newcommand{\sqwi}[1]{\sqrt{\left.\det w\right|_{#1}}}
\newcommand{\sqzwi}[1]{\sqrt{\left.\det \zw\right|_{#1}}}

\newcommand{\zwD}{{\mathring D}}

\newcommand{\npg}{\nabla_\hdw^\perp}
\newcommand{\npw}{\nabla_w^\perp}
\newcommand{\npzw}{\nabla_\zw^\perp}
\newcommand{\bmetric}{\mathbf{b}}
\newcommand{\dd}{\,\mathrm{d}}

\newcommand{\real}{\mathbb {R}}
\newcommand{\zahl}{\mathbb {Z}}
\newcommand{\eps}{\varepsilon}
\newcommand{\thet}{\vartheta}
\newcommand{\vphi}{\varphi}
\newcommand{\Rtrzy}{\mathcal{R}}
\newcommand{\Rdwa}{R}

\newcommand{\const}{\mathrm{const}}

\newcommand{\lapl}{\Delta}
\newcommand{\lapd}{\stackrel{2}{\lapl}\!}

\newcommand{\zg}{{\mathring g}}
\newcommand{\zw}{{\mathring w}}

\newcommand{\zPhi}{{\mathring \Phi}}
\newcommand{\zPsi}{{\mathring \Psi}}

\newcommand{\kw}{{k_w}}

\newcommand{\aw}{{a_w}}

\newcommand{\gm}{g_{m}}
\newcommand{\sqw}{\sqrt{\det w}}
\newcommand{\sqg}{\sqrt{\det g}}
\newcommand{\sqzw}{\sqrt{\det \zw}}

\newcommand{\hdw}{\mathbf{g}}
\newcommand{\hdzw}{\mathring\hdw}
\newcommand{\sqhw}{\sqrt{\det \hdw}}

\newcommand{\ee}{\end{equation}}
\newcommand{\nn}{\nonumber}
\newcommand{\eq}[1]{(\ref{#1})}
\newcommand{\be}{\begin{equation}}
\newcommand{\bel}[1]{\begin{equation}\label{#1}}
\newcommand{\arxiv}[1]{\url{http://arxiv.org/abs/#1}}

\begin{document}
\title{Energy-minimizing two black holes initial data}
\author{Jacek Jezierski\thanks{Partially supported by grant
 EPSRC: EP/D032091/1. Email: \texttt{Jacek.Jezierski@fuw.edu.pl}.}
\\ Katedra Metod Matematycznych Fizyki\\ Wydzia\l{} Fizyki UW\\
ul. Ho\.za 69, Warszawa, Poland
\\ \\ Jerzy Kijowski\thanks{Email:
\texttt{kijowski@cft.edu.pl}.}, Szymon
\L\c{e}ski\thanks{Partially supported by the Foundation
for Polish Science. Email: \texttt{szleski@cft.edu.pl}.}
\\ Centrum Fizyki Teoretycznej PAN\\
Al. Lotnik\'ow 32/46, Warszawa, Poland} \maketitle\vfill
\begin{abstract}
An attempt to construct the ``ground state'' vacuum initial data
for the gravitational field surrounding two black holes is
presented. The ground state is defined as the gravitational
initial data minimizing the ADM mass within the class of data for
which the masses of the holes and their distance are fixed. To
parameterize different geometric arrangements of the two holes
(and, therefore, their distance) we use an appropriately chosen
scale factor. A method for analyzing the variations of the ADM
mass and the masses (areas) of the horizons in terms of
gravitational degrees of freedom is proposed. The Misner initial
data are analyzed in this context: it is shown that they do not
minimize the ADM mass.
\end{abstract}\vfill
\pagebreak[4]

\section{Introduction}
Equations of motion of charged point particles in classical
electrodynamics can be derived from field equations within the
following approach: a generic state of the composed ``particles +
field'' system is treated as a perturbation of the ``ground
state'' of the field, uniquely determined by the positions of the
particles, see \cite{Dirac}, \cite{Haag}. In this classical
approach the ground state is defined {\em via} a (non-local in
time) decomposition of the actual field into the {\em retarded}
(or {\em advanced}) field and the remaining ``radiation field''.
Unfortunately, to decide what is {\em the retarded} or {\em the
advanced} field, the entire trajectory of the particle must be
known. Recently, it was shown that such a non-causal procedure may
be avoided and the ground state may be defined {\em via} a
conditional minimization of the (appropriately defined, ``already
renormalized'') total energy of the ``particles + field'' system,
with the positions and the charges of the particles being fixed
(see e.g.~\cite{K}, \cite{GKZ}, \cite {KP}). Mathematically, this
leads to a simple (elliptic) variational problem for the behaviour
of the field in a topologically non-trivial region of ${\mathbb R}^3$
(exterior of the particles), where the charges of the
particles provide the necessary boundary conditions. In this
context, the ground state of the many particle system may be
defined as the state of the field which contains ,,minimal amount
of the radiation field'', under condition that the charges and the
positions of the particles are fixed.

There is {\em a priori} no obstruction against applying a similar
idea to the problem of motion, and in particular to the two body
problem, in General Relativity theory. Here, we have to replace
point particles (elementary charges in electrodynamics) by black
holes (elementary masses in General Relativity theory) and to
consider perturbations of the field around a hypothetical ``ground
state'' of the two black holes system. Such a ground state could
be looked for as the state of the field, for which a conditional
minimum of the total energy of the ``black holes + gravitational
radiation'' system is achieved. Here, the total energy is equal to
the ADM mass at infinity. This way we are led to a variational
problem in a topologically non-trivial region of ${\mathbb R}^3$
(exterior of the two holes), where the variations must respect: 1)
the Gauss-Codazzi constraints for the field data, 2) the masses of
the holes (assigned to each of the horizons surrounding every hole
and playing role of the boundary conditions) and, finally: 3) the
fixed positions (distance) of the holes. The first two conditions
are technically complicated but the last one is even more
difficult to handle because it is ambiguous: the distance between the
two black holes (horizons?) may be measured in many nonequivalent
ways.

The idea of energy minimization is supported by the fact that, for
a single black hole, the initial data for the Schwarzschild metric
on the slice $t=\const.$ correspond, indeed, to the minimum of the
energy (ADM mass), with the area of the minimal surface
surrounding it being fixed (see e.g.~\cite{CQG2}, \cite{APP94},
\cite{MM}, \cite{GH-TI}). Thus, the Schwarzschild initial data may
be treated as a ``ground state'' of the one-body system. These
data have vanishing extrinsic curvature and the induced
three-metric is conformally flat. They lead to a static solution
of Einstein equations.

In electrodynamics, the ground state of the many body system is
always time symmetric. Indeed, any non-symmetricity means that the
magnetic field does not vanish and increases the energy of the
state. A similar argument (non-vanishing external curvature could
only increase the total energy of the state), together with the
example of the one-body problem (Schwarzschild) leads us to the
conclusion that the gravitational two-body ground state should be
looked for among time symmetric field configurations only. Such a
conjecture is supported by the fact that critical points for the
ADM mass correspond to the initial data generating stationary
vacuum space-times (see e.g.~\cite{Bartnik:2004ii},
\cite{Sudarsky:1992ty}).

Similarly as in electrodynamics, where there is no static solution
for the two particles system, in General Relativity there is no
static solution for two black holes (see \cite{Masood}). The
actual evolution of a hypothetical ``two black holes +
gravitational field'' system state will surely display interaction
between the ,,mechanical degrees of freedom'' (described by
positions and velocities of the holes) and the radiation degrees
of freedom. We very much hope that the perturbation techniques
based on the expansion with respect to the latter, in a
phase-space neighbourhood of the ground state, could provide a
good tool for the analysis of this evolution, at least in its
initial phase. In the final stage of the evolution, treating the
velocities on a perturbational level could become
non-satisfactory. Then, a ``dynamical ground state'' would be
necessary, where the time symmetry of the field configuration is
dropped and where the velocities of the holes are encoded in the
field boundary data on the horizons. This problem will be
discussed in the future.

In the present paper we assume, therefore, vanishing extrinsic
curvature of the field initial data in question, which also
implies vanishing of the linear and angular momentum. This
assumption simplifies the analysis of the constraints, vector
constraints becoming trivial. What remains is the analysis of the
scalar (Hamiltonian) constraint. On the other hand, we do not
restrict ourselves to conformally flat metrics. Such an
assumption, used by many authors, (see \cite{giu:cts},
\cite{mis:wic}, \cite{Brill-Lind}), freezes all gravitational
radiation degrees of freedom which are contained in the conformal
part of the three-metric and --- as shown in
Section~\ref{Misnerdata} --- is incompatible with the minimization
of the energy. Our strategy is to go beyond conformal flatness but
keeping the same topology of the exterior of the two black holes.
As a conclusion, we assume that initial data corresponding to the
ground state we are looking for are: 1) time symmetric (no
extrinsic curvature), 2) asymptotically flat, 3) the
three-dimensional region $\Sigma$ where the field data are living
is topologically $\Sigma = {\mathbb R}^3 \setminus {\cal B}_1
\setminus {\cal B}_2$ with two finite balls ${\cal B}_1$ and
${\cal B}_2$ being removed, and 4) the boundaries of these balls
are minimal surfaces.

Our goal is to construct the (momentarily static) two black holes
ground state as the vacuum initial data for gravitational field on
$\Sigma$ (i.e. outside of the two minimal surfaces) for which the
ADM mass (given by a boundary term at spatial infinity) is minimal
within the set of three-metric tensors respecting: 1) the given
masses of the holes and 2) their distance. Here, the following
problems arise:
\begin{itemize}
   \item How do we define the mass of each black hole? This can be done
in many ways, e.g. using quasi-local mass approach (see
\cite{Szabados}). Our choice is to assume that the mass
of a black hole corresponds to the area of the minimal
surface.
  \item How do we define the distance between black holes? Instead of
e.g.~geodesic distance, we propose to use a certain global
parameter defined at spatial infinity, which seems to parameterize
correctly the possible configurations of the two holes.
\end{itemize}

Our analysis of the scalar constraints enables us to
construct an integral identity which relates the variation
of the ADM mass with the changes of the three-metric in the
volume and its behaviour at the boundary. When applied to
the one black hole system, our procedure is strongly
related to the proof of the Penrose inequality
(\cite{PenIneq}).

The integro-differential equations which result from our
variational procedure are relatively complicated: there is little
hope to be able to solve them analytically and to obtain an
explicit formula for the gravitational ground state of the two
black holes system. We hope, however, that an appropriate,
numerical approximation will be useful for the description of the
complete two body problem. In this paper we test our methods,
applying them to simplified situations and show that this way we
easily reproduce classical results: the Minkowski and
Schwarzschild initial data are stationary points of the ADM mass
(in the latter case the mass of the horizon must be fixed {\em a
priori}). As a by-product of our method we show that the two black
hole Misner data do not describe the minimum of the ADM mass.

The paper is organized as follows: in Section \ref{sec:cdm} we
present our setup, proposed for description of the two-body data.
We use a ``2+1 foliation'' which mimics the bispherical system of
coordinates in the flat space ${\mathbb R}^3$ and we define
unconstrained degrees of freedom which parameterize the admissible
metrics. In the next Section we express the masses of the two
interacting black holes and the total (ADM) mass of the ``black
holes + gravitational field'' system in terms of the quantities
defined in Section \ref{sec:cdm}. In Section \ref{sec:lin} we
calculate variation of the ADM and the horizon masses in terms of
the variation of the metric degrees of freedom. First, the
calculation is performed for the conformally flat data and the
result is applied to the analysis of the Schwarzschild (Section
\ref{sec:ex}) and Misner (Section \ref{Misnerdata}) data. Section
\ref{sc:ogolnew} is the main part of the paper, where the above
analysis is generalized to a generic (not necessarily conformally
flat) case. In Section \ref{wzwcech} we argue, using symmetry
requirements, that the ground state of the two black hole system
has only one non-trivial degree of freedom. Then, we analyze in
detail this simplified (yet physically well-motivated) case.
Finally, equations for the ground state are derived in Section
\ref{sec:concl}. Technical details have been shifted to the
Appendices.

\section{Conformal decomposition of the metric}
\label{sec:cdm} Consider a three-metric $g$ (a part of an initial
data set for the gravitational field) on $\Sigma$, where $\Sigma$
is a three-dimensional manifold with boundary, which we obtain
removing two finite balls ${\cal B}_1$ and ${\cal B}_2$ from
${\mathbb R}^3 $. The boundary of $\Sigma$ consists of two
disjoint connected (spherical) components $H_i =
\partial {\cal B}_i$: $\partial\Sigma = H=H_1\cup H_2$.  We
assume $g$ to be a Riemannian metric on $\Sigma$, asymptotically
flat in the following sense: \be g=\Psi^4 \hdw\, ,\qquad \hdw =
\bmetric + \rho \,  , \ee where the conformal factor $\Psi$ is a
positive function on $\Sigma$, $\bmetric$ is a flat three-metric
and we impose appropriate decay conditions for the metric $g$,
namely $\rho=O(1/r^2)$, $\partial\rho=O(1/r^3)$, $\Psi=O(1)$ and
$\partial\Psi=O(1/r^2)$. We show in the sequel (cf.~Section
\ref{sec:concl} and Appendix \ref{ssc:btzero}) that these decay
conditions are compatible with the ground-state problem.

We are going to use on $\Sigma$ a ``bi-spherical'' system
of coordinates. More precisely, we parameterize the
one-point compactification of $\Sigma$ by the cylinder
\[
   \Sigma \cup \{ \infty \} \simeq S^2 \times I := W\, ,
\]
where $S^2$ is the topological two-sphere and $I$ is an interval
$[-a,b] \subset \real$.

On $W$ we introduce coordinates $x^i$, $i=1,2,3$, adapted to the
foliation, i.e. $\mu=x^3 \in [-a,b]$, and $x^A$, $A=1,2$, are
spherical coordinates $(\eta, \varphi )$ on each of the spheres
$\mu=\const$. We assume that $H_1$ ($H_2$, respectively)
correspond to the value $\mu=-a$ ($\mu=b$, respectively) and that
infinity corresponds to the north pole ($\eta = 0$) on the sphere
$\mu = 0$.

Such parameterizations are subject to a three-parameter (per point)
diffeomorphism group. To choose a specific one amongst them, three
gauge conditions have to be imposed. To fix coordinates
$(\eta,\varphi)$ on each sphere $\{ \mu = \mbox{const.}\}$, we
assume that the two-dimensional part of the metric $g$ (and,
therefore, also of the metric $\hdw$) is proportional to the
standard, unit, round metric on $S^2$:
\begin{equation}\label{round}
   \sigma_{AB}\dd x^A \dd x^B = \dd\eta^2 +\sin^2\eta \dd\vphi^2 \, .
\end{equation}
The above (two per point) gauge conditions give rise to the
further conformal decomposition of the metric $\hdw$:
$$ \hdw =h^2 w \, ,$$
where $w$ is a metric on the cylinder $W:=S^2\times I$, such that
its restriction to every leave coincides with \eqref{round}:
$w_{AB} = \sigma_{AB}$. The entire information about such a metric
is contained, therefore, in the following vector-density:
\begin{equation}\label{electric}
    D^k :=  \sqrt{\det g_{AB}} \ g^{3k} = \sigma w^{3k} \, ,
\end{equation}
where $w^{ij}$ and $g^{ij}$ are the inverse (contravariant)
metric tensors and $\sigma = \sqrt{\det \sigma_{AB}} = \sin
\eta $. The remaining gauge condition, fixing the
coordinate $\mu$ (and, therefore, leaves of the foliation
$\{\mu =  \mbox{const.}\}$), may be expressed in terms of a
differential equation imposed on the ``electric field''
$D$. In previous works \cite{CQG2}, \cite{APP94}, \cite{Grossmann}, \cite{PRD87},
conditions such as $\partial_k
D^k=0$ were successfully used in topologically different
arrangements. Most results of this paper do not rely on a
specific choice of the gauge. As will be seen in Section
\ref{wzwcech}, some results simplify considerably if we
assume vanishing of the mean external curvature of the
leaves $\{\mu =  \mbox{const.}\}$, calculated with respect
to the metric $w$ (i.e. condition $k_w=0$). The
applicability of this condition in a generic situation
needs further investigation.

Having at our disposal  the product of two conformal
factors: $\Psi$ and $h$, we are free to fix arbitrarily one
of them. We choose the standard value:
\[ h:= \frac{1}{\ch\mu-\cos\eta} \,  ,\]
which relates the flat, Euclidean three-metric $\bmetric$ on
$\real^3$, to the cylindrical metric $\zw=\dd\mu^2 +
\dd\eta^2 + \sin^2\eta \dd \vphi^2$ on $W$:
\begin{equation}\label{bdec}
    \bmetric = \dd x^2 + \dd y^2 + \dd z^2 =
  \frac{1}{(\ch\mu - \cos\eta)^2} [\dd\mu^2 +
  \dd\eta^2 + \sin^2\eta   \dd\vphi^2] = h^2 \zw \, ,
\end{equation}
{\em via} the bispherical coordinates $(\mu,\eta,\vphi)$:
\begin{eqnarray}
 x &=& \cos\vphi\frac{\sin\eta}{\ch\mu - \cos\eta}\, ,\label{x}\\
 y &=& \sin\vphi \frac{\sin\eta}{\ch\mu - \cos\eta}\, ,\\
 z &=& \frac{\sh\mu}{\ch\mu - \cos\eta}\, .\label{z}
\end{eqnarray}
The quantity $D$ takes in this case the following value:
\begin{equation}\label{D-zero}
    \zwD^3 = \sigma \, , \ \ \ \ \ \ \ \zwD^A = 0 \, .
\end{equation}

We shall use in parallel both of the two conformal
decompositions of the metric $g$ defined above. Putting
$$\Phi:=\Psi\sqrt h\, ,$$
we have: $$g=\Phi^4 w=\Psi^4 \hdw \, ,$$ and the following
(rescaling) transformation law for the Ricci scalar holds:
$$\Phi^4 \Rtrzy(g) = \Rtrzy(w) - \frac8\Phi \lapl_w \Phi\, ,$$
where by $\lapl_w$ we denote Laplace-Beltrami operator
associated with the metric $w$. The above formula implies
the following equations satisfied by the conformal factors
$\Phi$ and $\Psi$:
\bel{eqPhi}\sqw \Phi \left(\lapl_w - \frac{\Rtrzy(w)}{8}\right)
\Phi = -\frac18 \sqg \Rtrzy(g)\, ,\ee \bel{eqPsi} \sqhw \Psi
\left( \lapl_{\hdw}- \frac{\Rtrzy(\hdw)}{8} \right)\Psi =
-\frac18 \sqg \Rtrzy(g)\, .\ee For the special case $w=\zw$
and $\Phi = \sqrt h = 1/\sqrt{\ch\mu - \cos\eta}$ we have:
$g=\bmetric$, $\Rtrzy(\zw)=2$, $\Rtrzy(g)=0$ and equation
\eq{eqPhi} takes the form:
\begin{equation}\label{bez-delty}
   \left(\lapl_\zw-\frac 14\right)\sqrt{h} = 0\, .
\end{equation}
However, because of singularity of the function $\sqrt h$
at the point $\mu=0$, $\eta=0$ (i.e. at infinity), this
condition is fulfilled outside of this point, only. It is
easy to check that, globally, the following distributional
equation is satisfied, instead of \eqref{bez-delty}:
\bel{eqPhi2}
\left(\lapl_\zw-\frac 14\right)\sqrt{h} =
-4\pi\boldsymbol\delta_0\, .
\ee
This means that for any (smooth) test function $f \in
C_0^\infty (W)$ we have:
\bel{wyjdelty}
\int_W \Phi \left(\lapl_\zw f -\frac14 f\right)\sqzw =
-4\pi f(\mu=0, \eta=0)\, .
\ee
The Dirac delta distribution here reflects the fact that a
small sphere surrounding $\mu=0$, $\eta=0$ maps to a large
sphere in the metric $\bmetric$. In the limit,  point
$\mu=0$, $\eta=0$ corresponds to the sphere at infinity
$S_\infty$.

The two ends of this cylinder, obtained for $\mu \rightarrow \pm
\infty$, correspond to the points $(x,y,z)=(0,0,\pm1)$ in the
flat, Euclidean three-space. We can interpret these points as two test
bodies in the flat-metric limit. The distance between these two
points is standard and equals $2$. To allow arbitrary distances,
there are two possible methods: 1) to change transformation laws
\eqref{x}--\eqref{z}, or 2) to allow an extra multiplicative
factor $d \in \real_+$ in formula \eqref{bdec}. In this paper we
are going to use always the last option, which implies also an
extra multiplicative factor $\sqrt d$ at the right-hand side of
equation \eq{eqPhi2}. Then $g=d\cdot \bmetric$ and the distance
between endpoints $\mu=\infty$ and $\mu = -\infty$ equals $2d$.
The parameter $d$ may be thought of as a ``scale factor''. On the
other hand, the position of the interacting heavy bodies (black
holes) in the coordinate space will always be standardized.

In a generic (not necessarily flat) case, the scalar constraint
$\Rtrzy(g)=0$, together with geometric identities \eqref{eqPhi}
and \eqref{eqPsi}, imply the following equations:
\bel{rPsi}
\left(\lapl_{\hdw}-\frac {\Rtrzy(\hdw)}8\right)\Psi=0\, ,
\ee
\bel{rFi} \left(\lapl_w-\frac {\Rtrzy(w)}8\right)\Phi =
-4\pi\sqrt{d}\ \boldsymbol\delta_0\, ,
\ee
where, again, we have admitted an arbitrary scale factor
$\sqrt{d} = \left.\Psi\right|_{\infty}$ on the right-hand
side. (Note that $w\to\zw$ for $ \mu,\eta \to 0$. Hence, we
may use either $\sqw$ or $\sqzw$ in the integral
\eq{wyjdelty}.) In the sequel, we are going to interpret
$2d$ as a quantity parameterizing the distance between two
black holes, even beyond the limit of point-like bodies
in flat space. It was already proved \cite{Leski} that in
the limit $d \rightarrow \infty$ of the Misner data, the
value $2d$ gives, indeed, the exact distance between the
bodies.

For our purposes, however, it is sufficient to treat $d$ as a
scale factor: its rescaling by a multiplicative factor implies
rescaling of all the distances appearing in our problem. In any
specific geometric situation the value of $d$ is defined uniquely
once a specific gauge condition for the coordinate $\mu$ is
chosen.

In a generic case of two black holes we constrain  $\mu$ to
a finite interval $[-a,b]$. Given ``independent degrees of
freedom'' of the gravitational field, described by $D^k$
(i.e.~by the metric $w$), we want to retrieve the remaining
information about the physical metric $g$ from equation
\eq{rFi} on $W$. The Dirac delta and the scale factor $d$
in \eq{rFi} play role of ``boundary conditions at
infinity''. The remaining boundary conditions on the
spheres $H_1:=\{\mu=-a\}$ and $H_2:=\{\mu=b\}$ are implied
by the fact that we want these two-surfaces to be horizons.
We assume, therefore, that they are minimal in the ambient
three-metric $g$, i.e.~that their mean extrinsic curvature
$k$ vanishes: $k=0$. We use the following law of
transformation of $k$ under conformal rescaling:
\be \Phi^3 k = \Phi k_w - 4\npw\Phi\, ,\ee where $k_w$ is the
curvature calculated with respect to metric $w$, and $\npw$
is a projection (with respect to $w$) of the gradient on
the unit (again in metric $w$), outward normal vector. This
yields
\bel{wabrz} \left.k\right|_{H_i} = 0 \iff \left.\npw
\Phi\right|_{H_i} = \left. \frac{\Phi k_w}{4}\right|_{H_i}\, .\ee
In particular, if $k_w=0$, we get the Neumann boundary conditions
for  $\Phi$; for a particular, flat case $w=\zw$, we have
\bel{bFi} \left.k\right|_{H_i}=0 \iff\left.
\frac{\partial\Phi}{\partial\mu}\right|_{H_i} = 0\, .
\ee
For the conformal factor $\Psi$ these conditions may be
written as follows: (compare
\cite[equation~(36)]{coo:ida}):
\bel{bPsi} \Psi|_{\infty} = \sqrt d\, ,\qquad
\left.\left(\frac{\partial}{\partial\mu}\Psi + \Psi
\frac{\partial}{\partial\mu}(\log
\sqrt{h})\right)\right|_{H_i}=0\, .\ee

As an example of the two-body data fitting into the above
framework we may take the Misner metric \cite{mis:wic} or its
generalization \cite{mis:mig}, see \cite{Leski} for an explicit
formula. The metric is of the form:
$$g = \Psi_m^4\bmetric\, ,$$ and there are two minimal
surfaces surrounding the singularities of the conformal
factor. The minimal surfaces are metric spheres in the
conformal metric $\bmetric$, hence it is natural to use the
2+1 foliation given by bispherical coordinates; the minimal
surfaces correspond to $\mu = \pm \mu_0$ (or, to $\mu=-a$,
$\mu=b$ in case of non-equal masses). The conformal metric
$\bmetric$ is then rewritten in terms of the cylindric
metric \eqref{bdec}. On the other hand, the Brill and
Lindquist data, \cite{Brill-Lind}, are more difficult to
handle within this framework. This is due to the fact that
the minimal surfaces are not round spheres in the conformal
metric and, whence, we can not use the standard bispherical
coordinates to construct the foliation. However, it is
always possible to construct a topologically bispherical
foliation for these data, for which the metric $w$ is no
longer equal to $\zw$.

\section{ADM mass and masses of the horizons}
\label{sec:mass} For time-symmetric initial data the Hamiltonian
constraints reduces to $\Rtrzy(g)=0$. Equation \eq{eqPsi} can,
therefore, be written as: \bel{eqPsi2} -8\sqhw \nabla_\hdw
(\Psi\nabla_\hdw\Psi) = -8 \sqhw |\nabla_\hdw \Psi|^2 - \Psi^2
\sqhw \Rtrzy(\hdw)\, ,\ee where $|\nabla_\hdw \Psi|^2$ is the
length of the gradient of $\Psi$ with respect to $\hdw$. Let us
integrate formula \eq{eqPsi2} over the whole space $\Sigma$. The
complete divergence on the left-hand side of equation \eq{eqPsi2}
yields a boundary integral. Hence we get:
  \be \begin{split} -8
  \lim_{r\to\infty}\int_{S_r} \Psi\npg \Psi \sqhwi{S_r} - 8 \int_{H}
  \Psi \npg\Psi \sqhwi{H} \\ = - \int_{\Sigma} \left(8
  |\nabla_\hdw\Psi|^2 +
  \Psi^2\Rtrzy(\hdw)\right)\sqhw\, ,
  \end{split}\ee
where by $\npg$ we denote the projection (with respect to $\hdw$)
of the gradient on the unit, outward, normal vector (analogous to
$\npw$); $S_r$ are spheres of radius $r$ (in metric $\bmetric$).
By $\left.\hdw\right|_{S}$ we denote the pullback of the metric to
$S$, i.e. the induced metric. We are going to show in the sequel
that the first of the integrals yields the ADM mass:
\bel{madm} 16\pi m_{ADM} = -8
  \lim_{r\to\infty}\int_{S_r}
  \Psi \nabla_\hdw^\perp \Psi\sqhwi{S_r}\, .
\ee
For this purpose consider a three-metric $g$ of ADM mass $m$ and
let it satisfy the decay conditions we have imposed. Then, for
$d=1$, $\left.\Psi\right|_{\infty}=\sqrt d = 1$, the metric $g$
has the following form in terms of the asymptotic spherical
coordinates $(r,\theta,\vphi)$:
  \be g =
  \left(1+\frac{m}{2r}+O(\tfrac{1}{r^2})\right)^4 \left(\dd r^2 +
  r^2 \dd\thet^2 + r^2 \sin^2\thet\dd\vphi^2 +
  O(\tfrac{1}{r^2})\right)\, .
  \ee
Using the above form of the metric, formula \eq{madm} may be
checked by inspection. Rescaling of $r$ by an arbitrary factor
$d$, i.e. using the {\em Ansatz} $r = \widetilde{r} d$, leads to
an arbitrary value $\left.\Psi\right|_{\infty}=\sqrt d$ and to the
following, asymptotic form of the metric tensor:
\bel{schw} g
= \left(\sqrt d +\frac{m}{2\sqrt d \widetilde{r}}+
O(\tfrac{1}{\widetilde{r}^2})\right)^4 \left(\dd
\widetilde{r}^2 + \widetilde{r}^2 \dd\thet^2 +
\widetilde{r}^2 \sin^2\thet\dd\vphi^2 +
O(\tfrac{1}{\widetilde{r}^2})\right)\, .
\ee
Simple calculations show that equation  \eq{madm} still holds.

Using the relation between $\Phi$ and $\Psi$ we can express the
ADM mass in terms of $\Phi$:
\be  16\pi m_{ADM}=
  -8\lim_{\eps\to0}\int_{S_\eps} \left[\Phi\npw\Phi -
  \left(\frac{\Phi}{\sqrt{h}}\right)^2\sqrt{h} \npw\sqrt{h}\right]
  \sqwi{S_\eps}\, ,
\ee
where we subtract the renormalization term
$\sqrt{h}\nabla\sqrt{h}$ (corresponding to the flat metric)
from the term $\Phi\nabla\Phi$. We integrate over surfaces
$\eps=\const.$, where $\eps^2 = \mu^2 + \eta^2 $.

We define the mass of a black hole in terms of the area of the
minimal surface surrounding it\footnote{The energy $m_{H_i}$ plays
role of the lower bound in Penrose inequality (cf. \cite{PenIneq},
\cite{procesRP}) and never decreases according to the second law
of black hole physics (see e.g. \cite{Frol-Nov} or
\cite{Heusler}).}:
  \bel{mHdef}m_{H_i} =
  \sqrt{\frac{\int_{H_i}\lambda}{16\pi}}\, ,
  \ee
where by $\lambda$ we denote the two-dimensional volume
element on the leaves $\{ \mu = \const .\}$: $\lambda :=
\sqrt{\det g_{AB}}$.

Given a solution $\Phi$ of equation \eq{rFi} with boundary
conditions \eq{bFi} (or, equivalently, a solution $\Psi$ of
equation \eq{rPsi} with boundary conditions \eq{bPsi}) the
areas of the minimal surfaces (i.e.~the masses of black
holes) are given. We have, therefore, an indirect control
over these masses by an appropriate choice of $a$ and $b$.

\section{Variations of conformally flat initial data}
\label{sec:lin}

We are going to prove in the sequel that the ground state
of two interacting black holes cannot be described by
conformally flat data. The search of an appropriate ground
state must go, therefore, beyond conformal flatness. For
this purpose consider a perturbation of the conformally
flat metric. Denote
$$g = \Phi^4w\, ,\qquad\zg=\zPhi^4 \zw\, ,$$ and decompose the
degrees of freedom of the metric $w$ as deformations of the
degrees of freedom of $\zw$: $D^k := \zwD^k + \delta D^k$,
where $\zwD$ is given by equation  \eqref{D-zero}. Using formula
\begin{equation}\label{Delta}
    \lapl_g \Phi = \frac{1}{\sqrt g} \partial_i \left( g^{ij}
    \sqrt g \partial_j\Phi\right)
\end{equation}
we have the following linearization of the above quantity:
   \be \lapl_g \Phi = \lapl_\zg \delta \Phi +
   \left(1-\frac{1}{\sqrt \zg} \delta \sqrt g \right)\lapl_\zg \zPhi
  +  \frac{1}{\sqrt\zg}
  \partial_i \left(\delta\sqrt g \zg^{ij}\partial_j \zPhi +
  \sqrt\zg \delta g^{ij} \partial_j \zPhi\right)\, ,
  \ee
and, consequently, the following equation for $\delta\Phi$:
\begin{eqnarray}\label{rdphi}
\left(\lapl_\zw -\frac14\right) \delta\Phi & = & \frac18
\delta\Rtrzy\zPhi -\frac12 \frac{\delta D^3}{\sigma}
\lapl_\zw \zPhi \\
&& {}-\frac 1\sigma \partial_i \left[\left(\sigma \delta
w^{ij} - \frac 12 \delta D^3 \zw^{ij}\right)
\partial_j \zPhi\right]\nn\, .
\end{eqnarray}
We denote the right-hand side of this equation by $j[w]$:
\bel{jupr} j[w] = \frac18  \delta\Rtrzy(w)\zPhi -\frac12
\frac{\delta D^3}{\sigma} \lapl_{\zw} \zPhi -\frac 1\sigma
\partial_i \left[\left(\sigma \delta w^{ij} - \frac 12
\delta D^3 \zw^{ij}\right)
\partial_j \zPhi\right]\, .
\ee
Moreover, we keep the scale factor unchanged ($\delta d =
0$) and, whence, the right-hand side of the linearized
equation \eqref{rFi} vanishes. This is why the right-hand
side of \eqref{rdphi} contains no Dirac delta term. The
boundary condition for $\delta \Phi$ are such that $H_i$
remain minimal surfaces:
\bel{bdfi}
\left.\npzw \delta\Phi\right|_{H_i} = \left.\frac{\partial
(\delta \Phi)}{\partial \mu}\right|_{H_i} =0\, .
\ee

 The formula \eq{madm} for the ADM mass can also be
linearized around the metric $\hdzw=h^2\zw$:
\bel{dmadm}
16\pi \delta m_{ADM}= -8 \lim_{r\to\infty} \int_{S_r}
\left(\zPsi\npzw\delta\Psi +
\delta\Psi\npzw\zPsi\right)\sqhzwi{S_r}\, .
\ee
Because the contribution of the second term under the
integral vanishes, the formula may be rewritten as follows:
\bel{dmadmwar}
16\pi \delta m_{ADM}=
 -8 \lim_{r\to\infty}\int_{S_r}
\left(\zPsi\npzw\delta\Psi -
\delta\Psi\npzw\zPsi\right)\sqhzwi{S_r}\, .
\ee
In fact, we have $\delta\Psi\to 0$ at infinity (because the
scale factor is kept unchanged: $\delta d = 0$) and
$\npzw\zPsi = O(\tfrac {1}{r^2})$, which proves
\eqref{dmadmwar}. Rewriting it in terms of $\delta\Phi$ we
obtain:
\begin{equation}
16\pi\delta m_{ADM}  =
 -8 \lim_{\eps\to0}\int_{S_\eps} \left(\zPhi\npzw\delta\Phi -
\delta\Phi\npzw\zPhi\right)\sqzwi{S_\eps}\, .
\label{dmadm2}
\end{equation}

We are going to express the above variation of the ADM mass in
terms of variations $\delta D^k$ of gravitational degrees of
freedom. Observe that, due to elliptic equation \eq{rdphi}, the
variation $\delta\Phi$ of the conformal factor depends non-locally
upon variations $\delta D^k$. To handle this non-local dependence,
it is useful to rewrite the surface integral \eqref{dmadm2} in
term of a volume integral. Next, we shall transform the expression
in such way that the dependence upon $\delta D^k$ becomes
explicit. For this purpose we rewrite equations \eq{rdphi} and
\eq{eqPhi} in the following form:
\bel{eqdfi2}
\sqzw \zPhi\left(\lapl_\zw -\frac14\right) \delta\Phi =
\sqzw\zPhi j[w] \, ,\ee
\bel{eqfi2}\sqzw\delta\Phi
\left(\lapl_{\zw}-\frac14\right)\zPhi=0\, .
\ee
Subtracting these equations we get:
\be
\sqzw\zPhi j[w] = \sqzw\left(\zPhi\lapl_{\zw}\delta\Phi -
\delta\Phi\lapl_{\zw}\zPhi\right) =
\nabla_{\zw}\left(\zPhi\nabla_{\zw}\delta\Phi -
\delta\Phi\nabla_{\zw}\zPhi\right)\, .
\ee
If we multiply the above divergence by $-8$ and integrate it over
$W^*:=W\setminus \{(\mu=0, \eta=0)\}$ then the boundary term at
infinity reproduces precisely formula \eq{dmadm2}. Finally, we
obtain:
\begin{multline}\label{intj} 16\pi \delta m_{ADM} = 8
\int_H \left(\zPhi\npzw\delta\Phi -
\delta\Phi\npzw\zPhi\right)
\sqzwi{H} \\
{}- 8\int_{W^*} \zPhi \underbrace{\left(\lapl_\zw
-\frac14\right) \delta\Phi}_{j[w]}\sqzw\, .
\end{multline}
The first integral in \eq{intj} vanishes because of the
boundary conditions: \eqref{bFi} for  $\zPhi$ and
\eqref{bdfi} for $\delta\Phi$.
Consequently, we have:
\bel{dmphj} 16\pi \delta m_{ADM} = - 8\int_{W^*} \zPhi
j[w]\sqzw\, .
\ee
In the next step we rewrite variations $\delta\Rtrzy$ and
$\delta w^{ij}$ in $j[w]$ (see equation \eq{jupr}) in terms
of  $\delta D^k$. (Detailed calculations for $\delta
\Rtrzy$ have been shifted to Appendix \ref{app:rtrzy}.)
This way integral \eq{dmphj} may be rewritten as an
expression containing variations $\delta D^k$ and their
derivatives. In the last step we eliminate the latter using
integration by parts (boundary integral vanishes because we
assume that $\delta D^k$ vanish at the boundary). It is
convenient to formulate the final result in terms of the
following covector-valued, symmetric, bilinear form
$B_k(f,g)$:
\bel{BAupr}B_A(f,g) = -\frac14(fg)_{,3A}
+ 2f_{(,A}g_{,3)}\, ,
\ee
\bel{Btupr}B_3(f,g) = \frac18
\left(\lapd-1\right)(fg) - \frac12 \sigma^{AB}f_{,A}g_{,B}
+ \frac12 f_{,3}g_{,3}\, ,
\ee
where by $\lapd$ we denote the Laplace-Beltrami operator on
the unit sphere (in metric $\sigma$). Then the following
formula holds for an arbitrary function $f$:
\be \int_{W^*} fj[w] \sqrt{\zw} = \int_{W^*} B_k(f,\zPhi)
\delta D^k\, .
\ee
In particular, putting $f=\zPhi$, we may rewrite formula
\eqref{dmphj} as follows:
\be \label{fk1}16\pi\delta m_{ADM} =
 - 8\int_{W^*} B_k(\zPhi,\zPhi) \delta D^k\, .\ee

We want to restrict the above variation to the class of
those $\delta D^k$'s, for which the masses $m_{H_i}$, $i =
1,2$, of both our black holes remain unchanged. For this
purpose we are going to use the Lagrange multipliers method
as soon as we are able to express the variations $\delta
m_{H_i}$ in terms of appropriate volume integrals. We begin
with the following formula:
\be 16\pi\delta m_{H_i} = \frac{1}{2\mathring{m}_{H_i}}
\int_{H_i} \delta\lambda = \frac{1}{2\mathring{m}_{H_i}}
\int_{H_i} 4\zPhi^3\delta\Phi \sigma\, .
\ee
Denote by $\widehat{m}_{H_i}$ the (unique) function
satisfying equation
\bel{rownf}\left(\lapl_\zw-\frac14\right)\widehat{m}_{H_i}
=0
\ee
and boundary conditions
\bel{warf}\left. \npzw
\widehat{m}_{H_i} \right|_{H_j} =
\frac14\frac{\zPhi^3}{\mathring{m}_{H_i}}\  \delta_{ij} \, ,
\ee
where $\delta_{ij}$ is the Kronecker symbol. Due to
condition: $\npzw\delta\Phi|_{H_i}=0$, integration by parts
leads to the formula:
\be\begin{split}16\pi\delta m_{H_i} & = -8\int_{H_i}
\left(\widehat{m}_{H_i} \npzw \delta\Phi-\delta\Phi \npzw
\widehat{m}_{H_i}\right)\sigma \\ & = -8\int_{W^*}
\widehat{m}_{H_i}
\left(\lapl_\zw-\frac14\right)\delta\Phi\sigma \, ,
\end{split}
\ee
the last equality being true because the boundary term vanishes at
infinity. Hence, both quantities: $\delta m_{ADM}$ (cf. \eq{fk1})
and $\delta m_{H_i}$, may be expressed by similar volume integrals
of the form $ f\left(\lapl_\zw-\tfrac14\right)\delta\Phi$, where
$f$ is a solution to, respectively:
\be
  \left(\lapl_\zw-\frac14\right)f=\begin{cases}
  0 & $for\, $ m_{H_i} \cr
  -4\pi\boldsymbol\delta_{0}\cdot \sqrt d & $for\, $ m_{ADM}\end{cases}
\ee
with appropriately chosen boundary conditions. Such $f$ will be
called respectively \emph{the ADM mass increase factor} or
\emph{the horizon mass increase factor}. We already know that the
conformal factor is the ADM mass increase factor:
$\widehat{m}_{ADM} = \zPhi$. Hence, we have
\be 16\pi\delta m_{ADM} = -8\int_{W^*}\widehat{m}_{ADM}
\left(\lapl-\frac14\right) \delta\Phi \sqzw\, ,\ee
\be 16\pi\delta m_{H_i} =
-8\int_{W^*}\widehat{m}_{H_i} \left(\lapl-\frac14\right)
\delta\Phi \sqzw\, .
\ee

A field configuration minimizing the ADM mass within the
class of data with fixed masses of black holes must,
therefore, annihilate the following form:
\begin{eqnarray}
  16\pi \Big(\delta m_{ADM} - \sum_i \nu_i \delta m_{H_i}
  \Big)&=& -8\int_{W^*} \Big(B_k(\zPhi, \zPhi)-\sum_i \nu_i
  B_k(\widehat{m}_{H_i}, \zPhi)\Big)\delta D^k \nonumber \\
    &=& -8\int_{W^*} B_k\left(\Big(\zPhi -\sum_i \nu_i
    \widehat{m}_{H_i} \Big), \zPhi\right)
  \delta D^k \, , \label{52}
\end{eqnarray}
where $\nu_i$ are Lagrange multipliers. Observe that, moreover,
the above variation procedure respects the scale factor $d$ which
remains unchanged. We conclude that vanishing of the right-hand
side for an arbitrary variation $\delta D^k$ is necessary if our
field configuration has to realize the conditional minimum of the
total energy of the ``two black holes + gravitational field''
system, where the masses of the holes and their distance are
fixed\footnote{In fact, only two variations among the three
$\delta D^k$'s represent the change of the field configuration,
whereas the third one represents variation of a gauge condition --
at this point not yet fixed. }.

\section{Stability of the Schwarzschild initial data}
\label{sec:ex} To test our method we will apply the above
formulae to the flat metric $\bmetric$ and the
Schwarzschild metric $g_s$. The flat metric
$$
  \bmetric = d^2 h^2 \zw\,
$$
describes the field configuration surrounding two
``zero-mass black holes'', i.e.~two arbitrarily chosen
points of the flat space. We have: $B_k(\sqrt h, \sqrt
h)=0$. This corresponds in our approach to the (weak)
stability of the Minkowski initial data \cite{KlaNic}.

The Schwarzschild metric in the bispherical setting may be
written as follows (cf.~formula \eqref{schw}):
$$g_s = \Phi_s^4\zw = \left(\sqrt h \sqrt d+ \frac{m}{2\sqrt
d}\frac{\sqrt h}{\bar{r}}\right)^4 \zw \, .
$$
Here $m$ is the mass of the unique real horizon, whereas
$2d$ parameterizes its  ``distance'' from an arbitrarily
chosen fictitious ``zero-mass black hole''. The radius
$\bar r$ is a function of $(\mu, \eta, \vphi)$ and
parameters $m$ and $d$ (see Appendix \ref{app:Schw} for
details). In this case we have: $B_k(\Phi_s, \Phi_s)\neq0$,
because a generic deformation $\delta D^k$ changes the mass
of the hole. To calculate the right-hand side of \eqref{52}
we use an explicit expression for the $\widehat{m}_{H}$
factor:
$$\widehat{m}_H = \frac{m}{\sqrt d}\frac{\sqrt h}{\bar{r}}\, ,
$$
which may be easily verified. On the other hand, we have:
$$
 \Phi_s = \sqrt d\sqrt h + \frac{m}{2\sqrt d}\frac{\sqrt
h}{\bar{r}}\, ,
$$
and, consequently:
$$
\Phi_s = \sqrt d\sqrt h + \hat f\, ,\qquad
\widehat{m}_H = 2\hat f\, .
$$
The form $B_k(f,g)$ is bilinear and symmetric, hence we get
\begin{eqnarray*} \lefteqn{B_k(\Phi_s,\Phi_s) - \nu
B_k(\widehat{m}_H, \Phi_s) =} &&\\&& d \cdot B_k(\sqrt h,
\sqrt h) + (1-2\nu) B_k(\hat f,\hat f) + (2-2\nu)\sqrt d
B_k (\sqrt h,\hat f)\, ,
\end{eqnarray*}
where, similarly as in the flat case, we have $B_k(\sqrt h, \sqrt
h)=0$. It is also easy to check that $B_k(\hat f,\hat f)=0$.
Hence, the above quantity vanishes for $\nu = 1$. The
interpretation of this result is following: on a class of $D^k$'s
with appropriate asymptotics, a map $D^k \mapsto m_{ADM} [D^k]$
may be defined. Constraining the map to those $D^k$, for which
$m_H$ remains unchanged, we see that the linear part of $m_{ADM}
[D^k]$ remains unchanged, i.e. the Schwarzschild metric is a
stationary point of this map. In other words, a small deformation
of the Schwarzschild metric which does not change the area of the
horizon will not change the total ADM mass (cf.~\cite{CQG2},
\cite{Wald2}).

\section{Application to Misner data}
\label{Misnerdata}  The two-body data, which can be easily
analyzed using our method,  have been proposed by Misner
\cite{mis:wic}. The Misner metric $\gm = \Phi_m^4 \zw$ is
given by a conformal factor
\bel{Misphi}
\Phi_m=\sum_{n\in\zahl}
\frac{\sqrt{d}}{\sqrt{\ch(\mu+2n\mu_0) - \cos\eta}}\, ,
\ee
defined on $S^2\times [-\mu_0, \mu_0]$. The spheres
$\mu=\pm\mu_0$ are minimal surfaces of equal mass
$m=m(\mu_0, d)$. The above definition may be generalized to
the case of non-equal masses \cite{mis:mig}, the explicit
formulae for $\Phi$ are given in \cite{Leski}. According to
\eqref{D-zero} we have $\zwD^3 = \sigma$, $\zwD^A = 0$.

Our method of variation allows us to prove that, using a
small perturbation of the metric, one can decrease the ADM
mass of the above field configurations, without changing
the masses of both horizons and the scale factor $d$. 
(See \cite{doda, dodb} for previous results obtained using 
different techniques.) To
show this we use the fact that only two of three degrees of
freedom $D^k$ are independent: an additional gauge
condition fixing the foliation $\{ \mu = \const .\}$ may be
imposed. We choose the following gauge condition:
$$
  k_w=0\, .
$$
In this gauge we have
\bel{kwgage}
k_w= \left(\frac{\sigma D^{A}}{D^{3}} \right)_{,A} = 0 \, ,
\ee
hence, there is a function $\chi$ such that we have
($\eps^{AB}$ is the two-dimensional Levi-Civit\`{a} symbol):
\be
\delta \frac{\sigma D^{A}}{D^{3}}  = \eps^{AB}\chi_{,B}\, .
\ee
We have also
\be B_3 \delta D^3 + B_A \delta D^A = B_3 \delta D^3 +
\frac {\zwD^3}{\sigma}B_A \delta \left( \frac{\sigma
D^{A}}{D^{3}} \right) + B_A \frac{\zwD^A}{\zwD^3}\delta
D^3\,
\ee
and
\be -8\int_{W^*} B_k\delta D^k = -8\int_{W^*} \left[ \left( B_3
+ B_A \frac{\zwD^A}{\zwD^3}\right) \delta D^3 - \left(
\frac{\zwD^3}{\sigma} B_A \eps^{AB}\right)_{,B}
\chi\right]\, ,
\ee
where $B_k$ stands for $B_k(f, \zPhi)$. Now $\delta D^3 $ and
$\chi$ describe independent degrees of freedom and can be chosen
freely. The second term in the last integral (the response to
$\chi$) vanishes for any $f$ of interest here because of the axial
symmetry of the unperturbed metric (both $f$ and $g$ in
\eqref{BAupr} do not depend upon the variable $\vphi$). As
$\zwD^A=0$, the response to $\delta D^3$ is simply $B_3$. We want
to show that it is impossible to find such Lagrange multipliers
$\nu_1, \nu_2$ that $B_3(f,\Phi_m) =
B_3(\Phi_m-\nu_1\widehat{m}_{H_1}-\nu_2\widehat{m}_{H_2},\Phi_m)$
is identically zero, i.e. that $B_3(\Phi_m, \Phi_m)$,
$B_3(\widehat{m}_{H_1}, \Phi_m)$ and $B_3(\widehat{m}_{H_2},
\Phi_m)$ are linearly independent. We are unable to derive
analytic formulae for $\widehat m_{H_i}$, but it is a matter of
simple calculations to prove numerically this independence. For
this purpose we approximate $\Phi_m$ by truncating the series
defining its value (equation \eq{Misphi} in case of equal-mass data).
Then we calculate the boundary conditions for $\widehat m_{H_i}$
at $H_1$ and $H_2$ ($\mu=\pm\mu_0$ for equal-mass data), find
$\widehat m_{H_i}$ by solving equation $(\lapl-1/4)\widehat
m_{H_i} =0$, and finally calculate $B_3(\Phi_m, \Phi_m)$ and
$B_3(\widehat{m}_{H_i}, \Phi_m)$. The result shows that, indeed,
$B_3(\Phi_m, \Phi_m)$, $B_3(\widehat{m}_{H_1}, \Phi_m)$ and
$B_3(\widehat{m}_{H_2}, \Phi_m)$ are linearly independent. This
proves that the Misner data do not minimize the ADM mass.

\section{Variations of generic initial data}
\label{sc:ogolnew} The analysis of the Misner data
presented in the previous Section shows that to construct
the ground state we need to relax the condition $w=\zw$.
This statement is also supported by numerical analysis of
two-body data, see \cite{poi:one}. In this Section we
describe variations of generic initial data, not
necessarily conformally flat. For this purpose we consider
a generic perturbation $D^k \to D^k+\delta D^k$ of the
field data and calculate the linear part of the
corresponding perturbation of the conformal factor: $\Phi
\to \Phi+\delta\Phi$.

To shorten the formulae we will denote  $\sqw$ by $\sqrt
w$. The equations satisfied by $\Phi$ and $\delta\Phi$ on
$W^*$ take the form:
\bel{rnlinogolne} \left(\lapl_w -
\frac{\Rtrzy(w)}{8}\right) \delta\Phi = j[w]\, ,
\ee
\bel{rphiogolne}  \left(\lapl_w -
\frac{\Rtrzy(w)}{8}\right) \Phi = 0\, , \ee where
\bel{jfull} j[w] = \frac18 \Phi
\delta \Rtrzy + \frac{1}{\sqrt w} (\delta\sqrt{w}) \lapl_w
\Phi - \frac{1}{\sqrt w} \partial_i \left[ (\delta\sqrt w
w^{ij} + \sqrt w \delta w^{ij}) \partial_j\Phi\right]\, .
\ee
(Observe that this is a generic form of equation \eq{jupr}.)

We now calculate the coefficients  $B_k(f,\Phi)$, such
that
\be \int_{W^*} \sqrt w f \left(\lapl_w -
\frac{\Rtrzy(w)}{8}\right) \delta\Phi = \int_{W^*}
B_k(f,\Phi) \delta D^k\, .
\ee
We follow the procedure described in Section \ref{sec:lin},
with obvious modifications. While the expression for $B_k$
in case of $w=\zw$ were quite simple, in generic case they
get rather complicated. For example for $w=\zw$ we have
$\delta w^{AB}=0$, while in generic case
\be
  \delta w^{AB} = \frac{D^A \delta D^B}{\sigma D^3} +
  \frac{D^B \delta D^A}{\sigma D^3} -\frac{D^A D^B}{\sigma
  D^3}\frac{\delta D^3}{D^3}\, ,
\ee
the other variations get similarly complicated. Here, we give
final results for $B_k$ (the proofs has been shifted to Appendix
\ref{complicated}). It is interesting to compare the formulae
given below with equations \eq{BAupr} and \eq{Btupr} which hold
for $w=\zw$.

\begin{eqnarray}
B_3 & = & -\frac12 \frac{\sqrt w}{D^3} f \lapl_w \Phi -
\frac12 \frac{\sqrt w}{D^3} w^{ij} \Phi_{,i} f_{,j} +
\frac{\sqrt w}{\sigma}\Phi_{,3} f_{,3}- \frac {\sqrt
w}{\sigma(D^3)^2} D^A D^B
\Phi_{,A} f_{,B}\nn\\
&&{}+\frac{1}{8D^3} f\Phi
\partial_i\left[\sqrt w (k_w M_w^i +
a_w^i)\right]\nn\\
&& {}-\frac{1}{4} f\Phi k_{w,A} D^A
\frac1\sigma\left(\frac{\sqrt w}{\sigma}\right)^4 - \frac18
(f\Phi)_{,i}D^i\frac1\sigma\left(\frac{\sqrt w}{\sigma}
\right)^4 k_w\nn\\
&& {}+ \frac14 \frac{\sigma}{D^3} \frac{D^A}{D^3}
\left[(f\Phi)_{,i}D^i\frac{\sqrt w}{\sigma^2}\right]_{,A} +
\frac18\frac{\sqrt w}{D^3}\lapd (f\Phi) +
\frac{1}{8} f\Phi k_w^2 \frac{\sqrt w}{D^3}\nn\\
&& {}+\frac18 f\Phi k_{AB}k^{AB}\frac{\sqrt w}{D^3} - \frac
14 \left[\left(\frac{\sqrt w}{\sigma}\right)^2 f \Phi
k^{AB}D^3\right]_{,B}\frac{D^D}{(D^3)^2}\sigma_{AD}\nn\\
&& {}+\frac18 \left(\frac{\sqrt w}{\sigma}\right)^2 f\Phi
k^{AB}\left[ \left(\frac{D^D}{D^3}\sigma_{DA}\right)_{,B} +
\left(\mathrm{A}\leftrightarrow \mathrm{B} \right) \right]
\label{B3compl}
\end{eqnarray}
\begin{eqnarray}
B_A & = & 2\frac{\sqrt w}{\sigma}f_{(,3}\Phi_{,A)} +
2\frac{\sqrt w D^B}{\sigma D^3}f_{(,A}\Phi_{,B)} - \frac14
\left[ (f\Phi)_{,i}D^i \frac{\sqrt w}{\sigma^2}\right]_{,A}
\frac{\sigma}{D^3}\nn\\
&& {}-\frac{1}{4D^3}\left[ \left(\frac{\sqrt
w}{\sigma}\right)^2 f\Phi k^{BD}D^3\right]_{,B}\sigma_{AD}
- \frac14 \left(\frac{\sqrt w}{\sigma}\right)^2 f\Phi
k^{BD} \mathring{\Gamma}_{ABD}\nn\\
&& {}+ \frac14 \left(\frac{\sqrt w}{\sigma}\right)^2 f\Phi k_{w,A}
\label{BAcompl}
\end{eqnarray}
In the above formulae $k_{AB}$ denotes $k_{wAB}$ --- the extrinsic
curvature tensor of the $\{\mu=\const.\}$ leaves with respect to
$w$, $\mathring\Gamma_{ABC}$ are Christoffel symbols for the
metric $\sigma_{AB}$, $M_w^i$ is a normal vector, $M_w^i =
w^{3i}/\sqrt{w^{33}}$, and $\aw^i := \frac12 (w^{ij}-M_w^i M_w^j
)\partial_j \ln w^{33}$. Note also the relation $\displaystyle
\sqrt w = \sigma \sqrt{\frac{\sigma}{D^3}}$.

\section{Reduction to one degree of freedom}
\label{wzwcech} To analyze the expressions for $B_k$ obtained in
Section \ref{sc:ogolnew} we need first to make some simplifying
assumptions. In addition to the previously chosen gauge
$w_{AB}=\sigma_{AB}$ we impose the following gauge condition: $k_w
= 0$, i.e.~equation \eqref{kwgage}.
As shown in Section \ref{Misnerdata} the relevant integrand
is

\be \underbrace{\left( B_3 + B_A
\frac{D^A}{D^3}\right)}_{=:B}
 \delta D^3 - \left(
\frac{D^3}{\sigma} B_A \eps^{AB}\right)_{,B} \chi\, .
\ee
The degrees of freedom are $\delta D^3$ and $\chi$. The
response to $\delta D^3$, denoted by $B$, is a combination
of  $B_3$ and $B_A$, which is simpler than its components,
especially if rewritten in terms of $\displaystyle \kappa_{AB} = k_{wAB}
- \frac12 k_w\sigma_{AB}$:

\begin{eqnarray}
\label{surprisingly}
 B& = & -\frac12 \frac{\sqrt
w}{D^3} f \lapl_w \Phi + \frac12 \frac{\sqrt w}{D^3} w^{ij}
\Phi_{,i} f_{,j} - \frac {\sqrt w}{D^3} \sigma^{AB}
\Phi_{,A} f_{,B}\\
&&{}+\frac{1}{8D^3} f\Phi
\partial_A\left(\sqrt w a_w^A\right) +
\frac18\left(\frac{\sqrt w}{\sigma}\right)^3\lapd (f\Phi)
\nn\\
&& {}+\frac18 f\Phi \kappa_{AB}\kappa^{AB}\frac{\sqrt
w}{D^3} - \frac 14 \left[\left(\frac{\sqrt
w}{\sigma}\right)^2 f \Phi
\kappa^{AB}D^3\right]_{,B}\frac{D^D}{(D^3)^2}\sigma_{AD}
\nn\\
&& {}+\frac18 \left(\frac{\sqrt w}{\sigma}\right)^2 f\Phi
\kappa^{AB}\left[ \left(\frac{D^D}{D^3}\sigma_{DA}\right)_{,B} +
\left(\mathrm{A}\leftrightarrow \mathrm{B} \right) \right] \nn\\
\nn && {}-\frac{D^A}{4(D^3)^2}\left[ \left(\frac{\sqrt
w}{\sigma}\right)^2 f\Phi \kappa^{BD}D^3\right]_{,B}\sigma_{AD} -
\frac{D^A}{4D^3} \left(\frac{\sqrt w}{\sigma}\right)^2 f\Phi
\kappa^{BD} \mathring{\Gamma}_{ABD} \, .
\end{eqnarray}

Because of the cylindric symmetry of the problem, it is natural to
assume that its solution (the ground state of the two black holes
system) also respects this symmetry, i.e.~that the functions
$D^k$ do not depend upon the coordinate $\varphi$. In this case
gauge condition \eqref{kwgage} together with non-singularity of
the two-vector field $D^A$ on $S^2$ implies $D^\eta =0$. Moreover,
it is natural to assume the axial symmetry, i.e.~symmetry with
respect to transformation $\varphi \rightarrow - \varphi$. This
implies that $D^\vphi=0$ and $B_\vphi=0$. Hence, the response to
$\chi$ is zero:
\[
 \left(
 \frac{D^3}{\sigma} B_A \eps^{AB}\right)_{,B} = \left(
 \frac{D^3}{\sigma} B_\eta \eps^{\eta\vphi}\right)_{,\vphi}
 + \left(
 \frac{D^3}{\sigma} B_\vphi \eps^{\vphi\eta}\right)_{,\eta} = 0 \, .
\]

Denote
\be
u(\mu, \eta)= \frac{1}{\sqrt{w^{33}}}\, .
\ee
The function $u$ is now the only non-trivial degree of freedom
which we take into account. In this simplified situation we can
rewrite $\Rtrzy(w)$ as:
\bel{Rtrzysimple} \Rtrzy(w) = 2+ \frac{2}{\sqrt w}
\partial_A(\sqrt w a_w^A) = 2- \frac2u\lapd u\, .\ee
The response \eqref{surprisingly} to $\delta D^3$ reduces
to a surprisingly simple formula:
\bel{B3upr}
B_3(f,\Phi)  =  \frac{u^3}{2}  \left[ u^{-2} \Phi_{,3}
f_{,3} - \sigma^{AB} \Phi_{,A} f_{,B} +
\frac14\left(\lapd-1\right) (f\Phi)\right]\, ,
\ee
and, consequently, we have:
\begin{eqnarray}\label{warmu}
16 \pi \delta m_{ADM} & = & - 8 \int B_3(\Phi,\Phi)\delta
D^3 \\& = & - 8 \int \sigma \left[ u^{-2} (\Phi_{,3})^2 -
\Phi_{,A} \Phi^{,A} + \frac14\left(\lapd-1\right)
(\Phi^2)\right]\delta u\, .\nn
\end{eqnarray}
Equation \eqref{rFi} for $\Phi$ may now be written as
\be \partial_3
(u^{-1}\Phi_{,3}) + (u\Phi^{||A})_{||A} - \frac14\Phi\left(
1 - \lapd \right)u   = -4\pi\sqrt d\boldsymbol\delta_0 \, ,
\ee
where $_{||A}$ is a two-dimensional covariant derivative
with respect to $w_{AB} = \sigma_{AB}$. Equation
\eqref{rnlinogolne} for $\delta\Phi$ reduces now to
\begin{eqnarray}
\label{rnlinu}\lefteqn{\left(\lapl_w -
\frac{\Rtrzy(w)}{8}\right) \delta\Phi}&&\\
&&= \frac1u \bigg((u^{-2}\Phi_{,3}\delta u)_{,3} - \delta u
\lapd \Phi - \Phi^{,A}\delta u_{,A} +\frac14 \Phi\delta u -
\frac14 \Phi\lapd\delta u \bigg)\, \nn.
\end{eqnarray}

\section{Conclusions}
\label{sec:concl}

At this point we are able to fulfill the main goal of our paper:
to formulate necessary conditions for the ``ground state'' of
gravitational field around two black holes. Using the above
(physically well-motivated) reduction to a single degree of
freedom, described by the function $u$, the condition may be
formulated as follows: the conformal factor $\Phi$ satisfies
equation
\bel{concl1}
\partial_3 (u^{-1}\Phi_{,3}) + (u\Phi^{||A})_{||A} -
\frac14\Phi\left( 1 - \lapd \right)u   = -4\pi\sqrt
d\boldsymbol\delta_0 \, ,
\ee
on $W=S^2\times[-a,b]$ with Neumann boundary conditions at
$\mu = -a$ and $\mu = b$. Analogically,
$\widehat{m}_{H_i}$, $i=1, 2$, satisfy equations
\bel{concl2} \partial_3
(u^{-1}\widehat{m}_{H_i,3}) +
\left(u\widehat{m}_{H_i}^{||A}\right)_{||A} -
\frac14\widehat{m}_{H_i}\left( 1 - \lapd \right)u  = 0 \, ,
\ee
with boundary conditions \eqref{warf}. Let us define
\be\nn
f = \Phi - \sum_{i=1}^2 \nu_i\widehat{m}_{H_i}\, ,
\ee
where $\nu_1$, $\nu_2$ are Lagrange multipliers. To find
the ground state we look for such function $u(\mu,\eta)$
and numbers $\nu_1$, $\nu_2$ that $B_3(f,\Phi) = 0$, where
$B_3(f,\Phi)$ is given by equation  \eqref{B3upr}. This condition
reads:
\bel{concl4}
u^{-2} \Phi_{,3} f_{,3} - \sigma^{AB} \Phi_{,A} f_{,B} +
\frac14\left(\lapd-1\right) (f\Phi) = 0\, .
\ee
We have, therefore, four equations \eqref{concl1}--\eqref{concl4}
for four functions: $(u,\Phi, \widehat{m}_{H_i})$ and two Lagrange
multipliers. Once these equations are solved on $W$, the masses of
the holes can be read from the conformal factor $\Phi$; the
distance parameter is $2d$. The masses can be controlled
indirectly because we control the parameters $a$ and $b$. The
asymptotic analysis of equation $B_3(f,\Phi)=0$ proves that the
fall-off conditions for the metric $g$ we imposed at the very
beginning are satisfied for the ground state (see Appendix
\ref{ssc:btzero} for detailed calculations). This proves the
consistency of our approach.

Because of the high non-linearity of the problem, there is no
chance to solve it analytically. We do much hope that an
appropriate numerical analysis will allow to find solutions, which
would be a good starting point for a perturbational approach.

\appendix
\section{Linearization of the Ricci scalar}
\label{app:rtrzy}
 To derive a linear deformation of the Ricci scalar
 we start from the Gauss-Codazzi
identity:
\bel{GCI}
\Rtrzy(w) = \Rdwa(w) + \kw^2 - \kw_{AB}\kw^{AB} +
\frac{2}{\sqw}
\partial_i [\sqw(\kw M_w^i+\aw^i)]\, ,
\ee
where  $k_{wAB}$ is the extrinsic curvature tensor of the leaves
$\{\mu=\const.\}$, $k_w$ is its trace, $M_w^k$ is the normal
vector, $M_w^k = w^{3k}/\sqrt{w^{33}}$, and $\aw^i := \frac12
(w^{ij}-M_w^i M_w^j )\partial_j \ln w^{33}$. We have $w_{AB} =
\sigma_{AB}$, hence
\[
\Rdwa(w) = 2\, ,\]
\[\kw = \sigma^{AB}\kw_{AB} = -\frac{\sqrt{w^{33}}}{\sigma}\left(
\frac{\sigma w^{3k}}{w^{33}}\right)_{,k} \, .\] Inserting
the formulae $D^k=\zwD^k+\delta D^k$, we get, in notation
of Section \ref{sec:lin}, the following linear
approximations:
\[
\kw = -\sqrt{\frac{D^3}{\sigma}}\left[ \frac
{1}{D^3}D^k_{,k} +
\frac{D^k}{\sigma}\left(\frac{\sigma}{D^3}\right)_{,k}\right]
\approx -\frac1\sigma \delta D^A_{,A} \, ,
\]
\[
\kw^2 - \kw_{AB}\kw^{AB} \approx 0\, ,
\]
\[
M_w^A = O(\delta D^k)\, ,\quad \mathring{M}_w^3 = 1\, ,\quad
\aw^3=0\, ,\quad \aw^A \approx \frac 12 \sigma^{AB} \left(\frac{\delta
D^3}{\sigma}\right)_{,B}  \, .
\]
Incorporating these into the formula for $\Rtrzy$ we
finally get:
\be \label{Rtrzylin}\Rtrzy(w) -2 \approx
-\frac2\sigma (\delta D^A)_{,A3}  + \frac1\sigma \partial_A
\left[ \sigma\sigma^{AB} \left(\frac{\delta D^3}{\sigma}
\right)_{,B}\right]\, .\ee

\section{Schwarzschild metric -- bispherical foliation}
\label{app:Schw} The metric of the $t=0$ slice for the
Schwarzschild solution with the mass $m$ is (we fix the
value of the conformal factor $\Psi_s$ at infinity to be
$\sqrt d$):
$$g_s = \Psi_s^4 \bmetric = \Psi_s^4 (\dd x^2 + \dd y^2 + \dd z^2)\, ,
\qquad \Psi_s = \sqrt d+\frac{m}{2\sqrt d\bar r}\,,$$ where
$\bar r = \sqrt{x^2 + y^2 + (z-z_0)^2}$. We introduce the
bispherical coordinates $$x =
\frac{\sin\eta\cos\vphi}{\ch\mu - \cos\eta}\, ,\qquad y =
\frac{\sin\eta\sin\vphi}{\ch\mu - \cos\eta}\, ,\qquad z =
\frac{\sh\mu}{\ch\mu - \cos\eta}\, ,$$and choose such $z_0$
that the minimal surface $\displaystyle \bar r = \frac m{2d}$ coincides
with $\mu=-a$ sphere for some $a>0$. This leads to the
following formulae:
$$z_0 = - \sqrt{1+\frac{m^2}{4d^2}} = \frac{\ch a}{\sh a}\, ,
\qquad a =  \arsh \frac{2d}m
$$
$$ \bar r^2 = \left( \frac{\sh\mu}{\ch\mu-\cos\eta}
+ \sqrt{1+\frac{m^2}{4d^2}}
 \right)^2 + \frac{\sin^2 \eta}{(\ch\mu-\cos\eta)^2}
$$
$$\Phi_s = \Psi_s\sqrt h = \sqrt d\sqrt h + \frac m{2\sqrt d}
\frac{\sqrt h}{\bar r}
$$
$$\frac {\sqrt h}{ \bar r} = \left[\ch \mu + \cos \eta +
2\sqrt{1+\frac{m^2}{4d^2}} \sh\mu +
\left(1+\frac{m^2}{4d^2}\right)(\ch \mu -
\cos\eta)\right]^{-1/2}
$$

\section{Proof of formulae \eqref{B3compl} and \eqref{BAcompl}}
\label{complicated} In this Appendix we use notation introduced in
Section \ref{sc:ogolnew}. The first step in derivation of
\eqref{B3compl} and \eqref{BAcompl} is to express the linearized
term \eqref{jfull} (denoted by $j[w]$) in terms of $\delta D^k$.
It is a matter of straightforward calculation to see that
\begin{equation}
\frac{1}{\sqrt w} (\delta\sqrt{w})
\lapl_w \Phi = -\frac12 \frac{\delta D^3}{D^3}\lapl_w\Phi\, ,
\end{equation}
\begin{multline}
\label{bgen2} -\frac{1}{\sqrt w} \partial_i \left[
\left(\delta\sqrt w w^{ij} + \sqrt w \delta w^{ij}\right)
\partial_j\Phi\right] = \frac{1}{2\sqrt w}
\partial_i \left(\sqrt w
\frac{\delta D^3}{D^3}w^{ij}\partial_j\Phi\right)
\\ {}-\frac{1}{\sqrt w}\partial_3\left(\sqrt w
 \frac{\delta D^j}{\sigma}\partial_j\Phi\right)
 -\frac{1}{\sqrt w}\partial_A\left(\sqrt w \frac{\delta
D^A}{\sigma}\partial_3\Phi\right)
\\ {}-\frac{1}{\sqrt w}\partial_A
\left[\frac{\sqrt w}{\sigma D^3}\left( D^A\delta D^B +
D^B\delta D^A - D^AD^B \frac{\delta D^3}{D^3}\right)
\partial_B\Phi\right]\, .
\end{multline}
The remaining term, $\tfrac18 \Phi \delta \Rtrzy$, is more
complicated. As in Appendix \ref{app:rtrzy}, we start from the
Gauss-Codazzi identity (\ref{GCI}). The first term in $\delta
\Rtrzy$, namely $2k_w\delta k_w$, is calculated from
$$
k_w = -\frac1\sigma \sqrt{\frac{D^3}{\sigma}}\left(
\frac{\sigma D^k}{D^3}\right)_{,k}\,,
$$
which leads to
\be
\delta k_w = \frac12 \frac{\delta D^3}{D^3} k_w
-\frac{1}{\sqrt{w}} \left[\frac{\sigma
D^k_{,k}}{(D^3)^2}\delta D^3 - \frac{\sigma}{D^3}\delta
D^k_{,k} - \left(\frac{\sigma}{D^3}\right)_{,k}\delta D^k
+D^k \left( \frac{\sigma \delta D^3}{(D^3)^2} \right)
\right]\,.
\ee
To linearize the second term, $k_{wAB}k_w^{AB}$, we use the
following formula for $k_{wAB}$:
$$
k_{wAB} = \sqrt{\frac{\sigma}{D^3}} \Gamma^{3}_{AB}\,,
$$
where $\Gamma^{i}_{jk}$ are Christoffel symbols of the
metric $w$. The expression for $k_{wAB}$ in terms of $D^k$
is then
$$
k_{wAB} = \frac12 \sqrt{\frac{\sigma}{D^3}}
\left[ 2\frac{D^C}{\sigma}\mathring{\Gamma}_{CAB}
- \frac{D^3}{\sigma}\left(
\bigg(\frac{D^D}{D^3}\sigma_{DA}\bigg)_{,B}
+ \bigg(\frac{D^D}{D^3}\sigma_{DB}\bigg)_{,A}
\right)  \right]
$$
and the linearization can be easily calculated. The last
(third) term of $\Rtrzy$, the divergence, involves
$M_{w}^i$ and $a_w^i$. Once these are linearized, the
remaining calculations are straightforward. We have
$$
M_w^i = \frac{w^{3i}}{\sqrt{w^{33}}}\,, \qquad
a_w^3 = 0\,, \qquad a_w^A = -\frac12 \sigma^{AB}
\left(\frac{\sigma}{D^3}\right)_{,B}\frac{D^3}{\sigma}\,,
$$
hence
$$
\delta M_w^i = \frac{\sqrt{w}}{\sigma^2} \delta D^i
 - \frac12 \frac{\sqrt w D^i}{\sigma^2}\frac{\delta D^3}{D^3}
$$
and
$$
\delta a_w^A = \frac{\sigma^{AB}}{2\sigma}
\left[
\bigg(
\frac{\sigma}{D^3}\frac{\delta D^3}{D^3}
\bigg)_{,B} D^3
- \bigg(
\frac{\sigma}{D^3}
\bigg)_{,B} \delta D^3
\right]\, .
$$

We have, therefore, rewritten   $j[w]$  in terms of $\delta D^k$.
Once this is done, we may read $B_k(f,\Phi)$ from the integrand
$\sqrt w f j[w]$. For example, take the term
$$
-\int_{W^*}
f\partial_3\left(\sqrt w
 \frac{\delta D^j}{\sigma}\partial_j\Phi\right)
$$
arising from (\ref{bgen2}). Integration by parts yields
$$
\int_{W^*}
\sqrt w
 \frac{\delta D^j}{\sigma}\partial_j\Phi\partial_3 f  =
 \int_{W^*}\left(
\sqrt w \partial_3\Phi\partial_3 f \frac{\delta D^3}{\sigma}
+ \sqrt w \partial_A\Phi\partial_3 f \frac{\delta D^A}{\sigma}
\right)
 \, .
$$
This gives a contribution to $B_3(f,\Phi)$, equal $\frac{\sqrt
w}{\sigma}\partial_3\Phi\partial_3 f $, and a contribution to
$B_A(f,\Phi)$, equal $\frac{\sqrt
w}{\sigma}\partial_A\Phi\partial_3 f $. The contributions to
$B_k(f,\Phi)$ coming from all the other terms may be calculated in
exactly the same way. Finally, after rather tedious but simple
calculations, we obtain formulae \eqref{B3compl} and
\eqref{BAcompl}.

\section{Asymptotic behaviour of solutions of equation 
\mbox{$B_3(f,\Phi)=0$}}
\label{ssc:btzero} Equation \eq{concl4} for $B_3$ may be
analyzed asymptotically, that means near the point $(\mu=0,
\eta=0)$ (corresponding to spatial infinity).
We rewrite equation  \eq{concl4} as
\be
B_3  =  \frac{u}{2}  \left[ \Phi_{,3} f_{,3} -
u^2\sigma^{AB} \Phi_{,A} f_{,B} +
\frac{u^2}{4}\left(\lapd-1\right) (f\Phi)\right] = 0\, .\ee
We decompose $\Phi$, $f$ and $u$ as
\bel{rozklasymp}\Phi = \sqrt d \sqrt h + \tPhi\, ,\qquad
f = \sqrt d \sqrt h + \tf\, ,\qquad u = 1 + \tu\ee where $\tPhi$,
$\tf$, $\tu$ are bounded, and use $B_3(\sqrt h, \sqrt h)=0$.
Hence, vanishing of $B_3(f,\Phi)$ is equivalent asymptotically to
vanishing of the following quantity:
\begin{eqnarray}
\label{B3as}\frac2uB_3 & = & u^2 \left[ \frac14
(\lapd-1)(\tPhi\tf) - \tPhi^{,A}\tf_{,A}\right] + \sqrt d \sqrt
h_{,3} (\tPhi_{,3}+\tf_{,3}) \\ \nn && {}-\frac{u^2}{2} \sqrt d
\sqrt h^{,A} (\tPhi_{,A}+\tf_{,A}) + \tPhi_{,3} \tf_{,3} +
\frac{u^2}{4}\sqrt d \sqrt h (\lapd - 1)(\tPhi+\tf)
\\&&\nn {}-\frac14 d h(2\tu+\tu^2)
\\&&\nn{}+\frac{u^2}{4}\sqrt d
(\tPhi+\tf)\lapd\sqrt h  - \frac{d}{2} (2\tu+\tu^2)(\sqrt h_{,A}
\sqrt h^{,A} - \sqrt h \lapd \sqrt h)\, .
\end{eqnarray}
Let us expand this expression around $(\mu=0, \eta=0)$. We
introduce new coordinates $(\eps, \gamma)$:
\be \begin{split}
\mu & = -\eps\cos\gamma\, , \\
\eta & = \eps\sin\gamma\, .
\end{split} \ee
The asymptotic formulae
\bel{rozwh}\sqrt h = \sqrt2 \eps^{-1} +
O(\eps)\, ,\qquad \sqrt h_{,3} = +\sqrt2 \cos\gamma\eps^{-2}
+ O(1)\, ,\ee \be \sqrt h _{,\eta} = -\sqrt2 \sin\gamma
\eps^{-2} + O(1) \, , \ee
\be  \lapd\sqrt h = (\sqrt2 - 3\sqrt2 \cos^2\gamma)\eps^{-3} +
O(\eps^{-1}) = -\sqrt h_{,33}\,  \ee and the assumption
$\tu=O(\eps)$ allow us to extract the lowest power of
$\eps$, namely $\eps^{-3}$ from equation \eq{B3as}. The
only terms with $\eps^{-3}$ are the last two terms in
equation \eq{B3as} (the last line). Substituting asymptotic
expansions we get
\bel{B3asos} \frac2uB_3 = \frac{1}{\eps^3}\left[
\frac{\sqrt2}{4}\sqrt d (\tPhi+\tf) (1-3\cos^2\gamma) -
\frac{4d\tu}{\eps}\cos^2\gamma\right] +O(\eps^{-2})\, .\ee
By assumption $\tu$ is a differentiable function of $\mu$
and $\eta$, hence $\tu = \eps(C_1 \sin \gamma + C_2
\cos\gamma) + O(\eps^2)$, $C_1$, $C_2$ being constant. The
functions of $\gamma$ in \eq{B3asos} are linearly
independent and the necessary condition for asymptotic
vanishing of $B_3$ is
\bel{ubmale} \tu = O(\eps^2)\, ,\ee
\bel{warm0} \tPhi(\mu=0,\eta=0) +\tf(\mu=0,\eta=0) = 0\, .\ee
The condition \eq{ubmale} means that the metric $g$ is of
the form
\bel{asympg}g=\Psi^4 \left(\bmetric + O(r^{-2})\right)\, ,\ee
in accordance with the asymptotic conditions imposed in Section
\ref{sec:cdm}.

\end{document}